\documentclass[reprint,aps,showkeys,showpacs,pre]{revtex4-2}
\usepackage{bm}
\usepackage{docs}
\usepackage{physics}
\usepackage{graphicx,dcolumn}
\usepackage{amsfonts}
\usepackage{epsfig}
\usepackage{fancyhdr}
\usepackage{hhline}
\usepackage[titletoc]{appendix}
\usepackage[english]{babel}
\usepackage[colorlinks=true]{hyperref}
\usepackage[capitalize]{cleveref}
\usepackage{xcolor}
\usepackage{setspace}
\usepackage{natbib}
\usepackage{subcaption}
\usepackage{amsmath}
\usepackage{latexsym}
\usepackage{amssymb}
\usepackage{float}
\usepackage{textcomp}
\begin{document}
\title{Thermodynamic susceptibility as a measure of cooperative behavior in social dilemmas}
\author{Colin Benjamin}
\email{colin.nano@gmail.com}
\affiliation{School of Physical Sciences, National Institute of Science Education and Research, HBNI, Jatni-752050, India}
\author{Aditya Dash}
\email{aditya.dash@niser.ac.in}
\affiliation{School of Physical Sciences, National Institute of Science Education and Research, HBNI, Jatni-752050, India}
\begin{abstract}
	The emergence of cooperation in the thermodynamic limit of social dilemmas is an emerging field of research. While numerical approaches (using replicator dynamics) are dime a dozen, analytical approaches are rare. A particularly useful analytical approach is to utilize a mapping between the spin-1/2 Ising model in 1-D and the social dilemma game and calculate the magnetization, which is the net difference between the fraction of cooperators and defectors in a social dilemma. In this paper, we look at the susceptibility, which probes the net change in the fraction of players adopting a certain strategy, for both classical and quantum social dilemmas. The reason being, in statistical mechanics problems, the thermodynamic susceptibility as compared to magnetization is a more sensitive probe for microscopic behavior, e.g., observing small changes in a population adopting a certain strategy. In this paper, we find the thermodynamic susceptibility for reward, sucker's payoff and temptation in classical Prisoner's Dilemma to be positive, implying that the turnover from defect to cooperate is greater than vice-versa, although the Nash Equilibrium for the two-player game is to defect. In classical Hawk-Dove game, the thermodynamic susceptibility for resource suggests that the number of players switching to Hawk from Dove strategy is dominant. Entanglement in Quantum Prisoner's Dilemma (QPD) has a non-trivial role in determining the behavior of thermodynamic susceptibility. At maximal entanglement, we find that sucker's payoff and temptation increase the number of players switching to defect. In the zero-temperature limit, we find that there are two second-order phase transitions in the game, marked by a divergence in the susceptibility. This behavior is similar to that seen in Type-II superconductors wherein also two second-order phase transitions are seen.
\end{abstract}
\maketitle
{\bf This work aims to define and interpret the role of thermodynamic susceptibility in social dilemmas in the infinite player limit. It relies on a mapping of the social dilemma game to the spin-1/2 Ising model. Magnetization of the game provides the difference between fraction of cooperators and defectors in the game. Susceptibility, on the other hand, provides the difference between the rate of change in the fraction of cooperators or defectors in response to a change in payoffs. In this paper, we calculate the thermodynamic susceptibilities for the classical games of Prisoner's Dilemma, Hawk-Dove and the quantum game of Quantum Prisoner's Dilemma. Analogous to magnetic susceptibility, the susceptibilities for the game are much more sensitive in detecting changes in fraction of cooperators or defectors in response to change in payoffs. We identify phase transitions in social dilemmas by looking at susceptibility plots. Temperature in a social dilemma plays a role so as to randomize the strategy selection by the players. Zero temperature corresponds to zero randomness while infinite temperature corresponds to complete randomness. Our main results are that the thermodynamic susceptibility for reward, sucker's payoff and temptation in classical Prisoner's Dilemma are positive, implying that the turnover from defect to cooperate is greater than vice-versa, although the Nash Equilibrium for the two-player game is to defect. In the case of Hawk-Dove game, we find that change in resource value increases the number of players switching to Hawk, while change in injury cost aids in the change in players to Dove. In the case of quantum prisoner's dilemma in thermodynamic limit, we find that entanglement plays a crucial role. At maximal entanglement, we find that sucker's payoff and temptation both promote the transition to defect in the game, while reward and punishment do not affect the  transition in the game. In the zero temperature limit, we find that there are two second-order phase transitions in the game, namely from an all classical phase to a mixed phase where the number of classical and quantum strategy players are equal, and from the mixed phase to a quantum phase, which are clearly identified by the divergence in the susceptibility.}

\section{Introduction}
Social dilemma games involve interactions between intelligent rational decision makers. Each participant (player) in the game attempts to maximize his/her own payoffs, which may lead to conflict, although on many occasions, cooperation may be more rewarding. The major aim in any social dilemma is to obtain the Nash equilibrium, a set of strategies to be selected by each player, so as to avail least loss or maximum gain for all players. In addition to Nash Equilibrium, there may also exist Pareto optimal strategies which could provide a better outcome. Such a situation is best represented by the classic Prisoner's dilemma, in which the Nash equilibrium for players is to defect, but they will receive a better payoff if they cooperate among themselves, which is the Pareto optimal strategy \cite{Devos2016}.

There are many real life examples where number of players involved in the game may be very large. For example, decision making process for a country involves accounting of choices made by each citizen which may number in millions. In these situations, looking at social dilemmas in the thermodynamic limit makes sense. A method for modeling social dilemmas in the thermodynamic limit by establishing a one-to-one correspondence between the payoffs of a general bi-matrix symmetric game with an appropriate exactly solvable statistical 1D Ising) has been attempted before\cite{Sarkar2019}. The mapping provides the equivalent of the "J" and "h" parameters for the social dilemma in terms of payoff matrix. Further, this mapping enables an analytic expression for the difference between  number of cooperators and defectors, which is a welcome alternative to the numerical experiments mostly done to approach the thermodynamic limit of social dilemma games. These "J" and "h" parameters are then substituted in the thermodynamic functions like magnetization and susceptibility to generate the equivalent magnetization or susceptibility for social dilemmas in  thermodynamic or infinite player limit. Magnetization in Ising model, defined as the difference between fraction of spin up ($\uparrow$) and spin down ($\downarrow$) sites, is then for a social dilemma defined as the difference between fraction of players opting to cooperate versus fraction of players opting to defect.
Apart from magnetization, there are other thermodynamic functions of interest that can be calculated from the Ising Model. One such function is the susceptibility, which provides the response of magnetization to a change in the external magnetic field. 

In magnetic systems, susceptibility provides a much more sensitive way to measure small changes in magnetic moment of a system at high external fields \cite{Martien}. In this paper we look at the social dilemma equivalent of the susceptibilities 
which provides us with  fraction of players switching between strategies and its implications for infinite player or thermodynamic limit of Prisoner's Dilemma, Hawk-Dove Game and Quantum Prisoner's Dilemma(QPD). The susceptibilities in social dilemmas are defined with respect to each of the payoffs. In our work, we find that susceptibilities for reward, sucker's payoff and temptation in classical Prisoner's Dilemma are positive, indicating that the number of players changing from defect to cooperate increases as a function of these payoffs. In classical Hawk-Dove game, increasing resource leads to increase in the turnover of players to Hawk while increasing injury cost leads to increase in turnover of players to Dove. For QPD, at partial entanglement, reward aids in the  transition to quantum strategy while punishment aids in the change to defect. At maximal entanglement, sucker's payoff and temptation aid in increasing the switch to defect. 
In the zero temperature limit, we find that the QPD has two second-order phase transitions at entanglement values of $\gamma_1$ and $\gamma_2$, which mark the phase transitions from a classical phase to a random phase, and from a random phase to a quantum phase. These can be easily identified by the divergence in the susceptibility at two entanglement values of $\gamma_1$ and $\gamma_2$. At finite temperatures, the random phase disappears and a single phase transition at $\gamma_0$ occurs. The phase diagram of QPD in thermodynamic limit is akin to the phase diagram of a type-II superconductor\cite{tinkham_1996,kittel_2005}. 

This paper has the following layout. First, we introduce the spin-1/2 Ising model and derive the expression for magnetic susceptibility. Then we map Ising model to a generic social dilemma game and then derive expressions for the susceptibilities for each payoff. We then interpret the game susceptibility in thermodynamic limit of classical Prisoner's Dilemma, classical Hawk-Dove and Quantum Prisoner's Dilemma. Finally, we end with conclusion.
\section{Spin-1/2 Ising model and Mapping to a general social dilemma game}
\subsection{Spin-1/2 Ising model in 1D}
The 1D Ising model consists of half-integer spins which can take values $\pm 1$. Spins only interact with their nearest neighbors via coupling ($J$) and all spins are subject to an uniform external magnetic field ($h$). The model is described by a Hamiltonian as- 
\begin{eqnarray}
    H = -J\sum_{i=1}^N S_i S_{i+1} - h \sum_{i = 1}^N S_i , \label{IsingHamil}
\end{eqnarray}
where $S_i$ is the spin at site $i$ and $N$ is the number of spins in the chain.
The corresponding partition function for 1-D Ising Hamiltonian in (\ref{IsingHamil}) is 
\begin{equation}
Z = \sum_{S_1,S_2,...S_N} e^{\beta (J \sum_{i=1}^{N} S_i S_{i+1} + h \sum_{i=1}^{N} (S_i+S_{i+1})/2)},
\end{equation}
with $\beta = \frac{1}{k_B T}$ representing inverse temperature and $k_B$ is Boltzmann constant while $T$ is the temperature. To evaluate sum over all spins in the partition function, we utilize transfer matrix method \cite{Kramers1941}. A detailed explanation for same is given in \cite{Landi,Baxter1982}.
A transfer matrix $V$ can be defined with its elements as
\begin{equation}
	V(S_i,S_{i+1}) = e^{\beta(J S_i S_{i+1} + h (S_i + S_{i+1})/2)}.
\end{equation}
The transfer matrix $V$ for a two spin Ising system can then be written as
\begin{eqnarray}
V &=& \begin{bmatrix} 
V(1,1) & V(1,-1) \\
V(-1,1) & V(-1,-1) \\ 
\end{bmatrix} = \begin{bmatrix} 
e^{\beta (J + h)} & e^{-\beta J} \\
e^{-\beta J} & e^{\beta (J - h)}\\ 
\end{bmatrix}
\label{transfermatrix}
\end{eqnarray}
The full partition function in terms of transfer matrix elements $V(S_i,S_{i+1})$ can then be calculated for $N$-spin case as-  
\begin{equation}
Z = \sum_{S_1,...,S_N} \prod_{i=1}^{N} V(S_i,S_{i+1})
\end{equation}
Here, we assume that model has periodic boundary conditions, i.e. $S_{N+1} = S_1$. 
The transfer matrix $V$ has the property\cite{Landi},
\begin{eqnarray}
\sum_{S_2} V(S_1,S_2) V(S_2,S_3) &=& V^2(S_1,S_3), \label{transprop1}
\end{eqnarray}
We will utilize the eigenvalues of $V$ to compute magnetization and subsequently susceptibility of Ising model. Eigenvalues of $V$ from Eq.~(\ref{transfermatrix}) are  
\begin{equation}
\lambda_{\pm} = e^{\beta J} \left(\cosh(\beta h) \pm \sqrt{\sinh^2(\beta h) + e^{-4 \beta J}}\right).
\end{equation}
One can see that condition $\lambda_+ > \lambda_-$ always holds.
Using Eq.~(\ref{transprop1}), we evaluate the partition function by summing over all spins. Thus,
\begin{equation}
	Z = \sum_{S_1} V^N(S_1,S_1) = \Tr(V^N).
\end{equation}
By using properties of the trace of a matrix, partition function can be written in terms of eigenvalues as
\begin{eqnarray}
Z &=& \lambda_+^N + \lambda_-^N = \lambda_+^N \left( 1 + \left(\frac{\lambda_-}{\lambda_+}\right)^N\right). \label{partition_eigenvalues}
\label{eq:ZN}
\end{eqnarray}
Second term of Eq.~(\ref{eq:ZN}) vanishes in the thermodynamic limit, i.e., $N\rightarrow\infty$. This gives
\begin{eqnarray}
Z &=& \lambda_+^N. \label{partition_func}
\end{eqnarray}
\begin{figure*}
	\centering
	\includegraphics[width=0.80\linewidth]{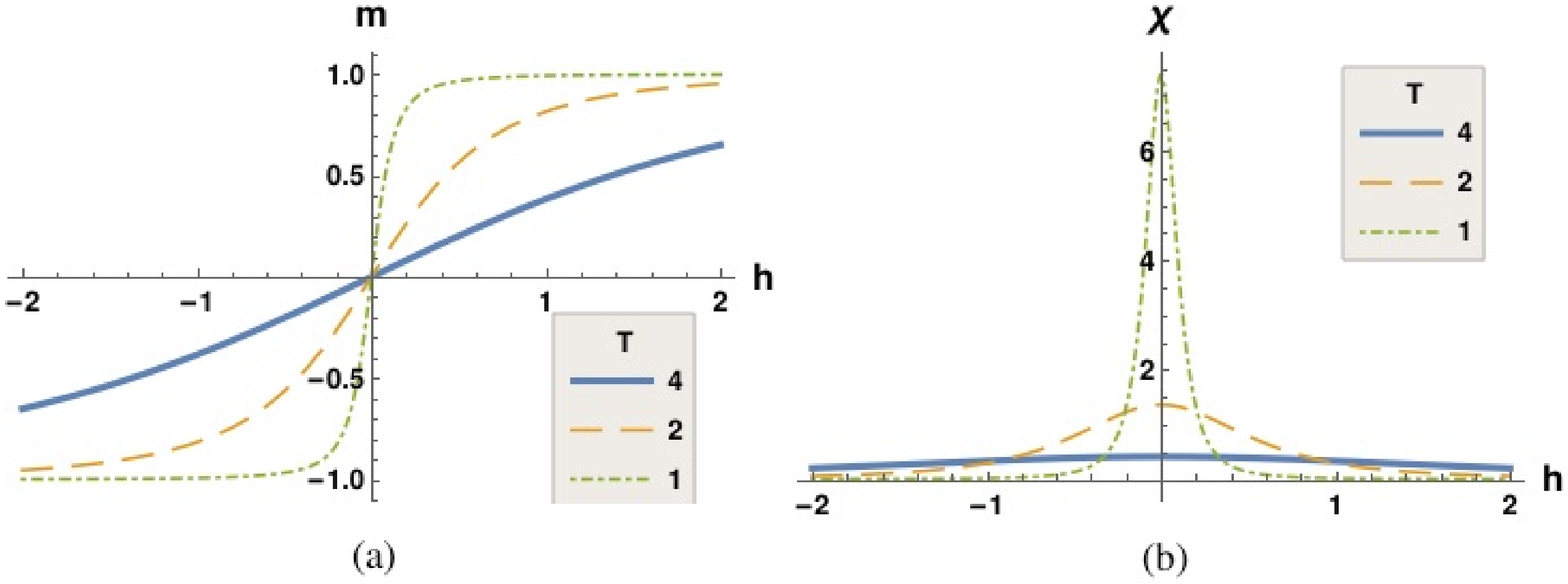}
\caption{Plot of (a) magnetization($m$) and (b) susceptibility($\chi$) for 1D Ising model as function of external magnetic field for different temperatures with $J=1$ and  $T = (\beta k_B)^{-1}$.\label{ising_plots}} 
\end{figure*}
To derive magnetization of Ising model, we begin with the expression of free energy per spin ($F$) for Ising model
\begin{equation}
	F = -\frac{1}{\beta N} \ln Z.
\end{equation}
In thermodynamic limit ($N \rightarrow \infty$), the free energy simplifies to 
\begin{equation}
	F = -\frac{1}{\beta} \ln \lambda_+.
\end{equation}
The average net magnetization is then
\begin{eqnarray}
	m&=&-\pdv{F}{h} = \frac{\sinh(\beta h)}{\sqrt{\sinh^2(\beta h) + e^{-4 \beta J}}}.\label{isingmag}
\end{eqnarray}
Magnetization provides the difference between fraction of up and down spins in Ising chain, i.e., $m = n_\uparrow - n_\downarrow$ and is shown in Fig.~1(a). Susceptibility $\chi$ is then defined as the derivative of magnetization in (\ref{isingmag}) with respect to the external magnetic field $h$, or 
\begin{equation}
    \chi = \pdv{m}{h}
\end{equation}
Susceptibility is a response function which provides the turnover of up spins or down spins as function of external magnetic field. Since sum of fraction of up and down spins in the chain is $n_\uparrow + n_\downarrow = 1$, we have
\begin{equation}
    \chi = 2 \pdv{n_\uparrow}{h},
\end{equation}
where $n_\uparrow$ is the fraction of up spins in chain. We have plotted both magnetization and susceptibility of the 1D Ising model in Fig.~1.
\subsection{Thermodynamic limit of social dilemma games \label{gengamedef}}
Consider a general bi-matrix symmetric game $G$ given as:
\begin{eqnarray}\label{gendoublepayoffmatrix}
G&=&\left(\begin{array}{c||c|c}
& s_1 & s_2 \\ \hhline {=#=|=} 
s_1 & a,a' & b,b' \\ \hline 
s_2 & c,c' & d,d'
\end{array}\right),
\end{eqnarray}
where $G(s_i,s_j)$ is the payoff function with $a, b, c, d$ as row player's payoffs  and $a', b', c', d'$ are column player's payoffs and $s_1,s_2$ are the strategies available to players.
To make a one-to-one correspondence of the payoffs with Ising model we require a set of transformations to payoffs. The transformations of payoffs are as follows \cite{Sarkar2019}:
\begin{eqnarray}\label{transformationpayoff}
G &=& \left(\begin{array}{c||c|c} 
& s_1 & s_2 \\\hhline{=#=|=} 
s_1 & a+\lambda,a'+\lambda' & b+\mu, b'+\lambda'\\ \hline
s_2 & c+\lambda,c'+\mu' & d+\mu, d'+\mu'
\end{array}\right).
\end{eqnarray} 
where $\lambda,\lambda',\mu,\mu'$ are the transformations to payoffs. These transformations do not alter the Nash equilibrium of game (See appendix of \cite{Sarkar2018,Sarkar2018b,Devos2016} for a general proof). Choosing the transformations as $\lambda=-\frac{a+c}{2},\lambda'=-\frac{a'+b'}{2}$ and $\mu=-\frac{b+d}{2},\mu'=-\frac{c'+d'}{2}$ and since game is symmetric, payoff matrix in (\cref{transformationpayoff}) is
\begin{eqnarray}\label{transformedpayoff}
G&=&\left(\begin{array}{c||c|c} 
& s_1 & s_2 \\\hhline{=#=|=} 
s_1 & \frac{a-c}{2},\frac{a-c}{2} & \frac{b-d}{2},\frac{c-a}{2} \\ \hline
s_2 & \frac{c-a}{2},\frac{b-d}{2} & \frac{d-b}{2},\frac{d-b}{2}
\end{array}\right).
\end{eqnarray} 
Next, we map the two player, two strategy social dilemma game to a two spin Ising model. The Ising Hamiltonian with two spins is 
\begin{equation}
    H = -J S_2 S_1 -J S_1 S_2 - h (S_1 + S_2) = E_1 + E_2.
\end{equation}
$S_1, S_2$ are spins at site 1 and 2 respectively. The energy at those two sites are then
\begin{equation}
    E_1 = -J S_1 S_2 - h S_1 \;\;\;\;;\;\;\;\; E_2 = -J S_2 S_1 - h S_2
\end{equation}
In order to map the 1D Ising model to the game as per recipe given in\cite{Galam2010, Sarkar2018,Sarkar2018b}, it is important to note that in Ising model, we have to minimize energy to obtain the equilibrium, in game theory, in contrast, Nash equilibrium is obtained by maximizing the payoffs. Hence, to map these two models, we consider the negative of  energy, i.e., $"-E"$, where $E$ is total energy in the Ising model. The energies of the spin configurations in two spin Ising model are written in matrix form as-
\begin{eqnarray}
-E &=& \left(\begin{array}{c||c|c}
& S_2=+1 & S_2=-1  \\\hhline{=#=|=} 
S_1=+1 & J+h,J+h & -J+h,J-h \\ \hline
S_1=-1 & J-h,-J+h & -J-h,-J-h 
\end{array}\right). \label{energysitematrix}
\end{eqnarray}
As in\cite{Galam2010}, equating (\ref{transformedpayoff}) with (\ref{energysitematrix}), we get $J+h = (a-c)/2$, $J -h = (c-a)/2$, $-J + h = (b-d)/2$ and $-J-h = (d-b)/2$ and we obtain the relation between $J,h$ and $a,b,c,d$ as -
\begin{eqnarray}
    J &=& \frac{a-b+d-c}{4} \;\;\;\; ; \;\;\mbox{and }\;\ h = \frac{a-c+b-d}{4}. \label{genrelation}
\end{eqnarray}
Utilizing this mapping, we obtain the game magnetization $m_g$, i.e., the difference between fraction of players who have selected strategy $s_1$ and strategy $s_2$, analogous to that of magnetization in Ising model, for a general social dilemma game in thermodynamic or infinite player limit as
\begin{eqnarray}
	m_g &=& \frac{\sinh \left(\frac{a+b-c-d}{4 T}\right)}{\sqrt{e^{\frac{-a+b+c-d}{T}}+\sinh ^2\left(\frac{a+b-c-d}{4 T}\right)}}, \label{genmag}
\end{eqnarray}
where $T = (k_B \beta)^{-1}$ is game temperature with $k_B$ being Boltzmann constant and $\beta$ being the inverse temperature. 
An important point to note is that the mapping discussed here is between the Nash equilibrium of the payoff matrix of a $2 \times 2$ symmetric non-zero sum game (\ref{transformedpayoff}) to a two spin Ising energy matrix(\ref{energysitematrix}). A successful mapping would imply that the Ising two spin energy matrix will now be called as an Ising two spin payoff matrix. Thus we are not mapping the eigenvalues of the payoff matrix(\ref{transformedpayoff}) to the eigenvalues of the Ising two spin energy matrix(\ref{energysitematrix}) but rather the Nash equilibrium of the payoff matrix to the Nash equilibrium of the two spin Ising payoff matrix. In a separate work~\cite{Sarkar2018b}, one of us, has also shown that an approach which purports to map eigenvalues of payoff matrix to   eigenvalues of the Ising two spin model leads to incorrect results in the appropriate limits, since eigenvalues of payoff matrix do not correspond to anything of relevance and importantly they do not have any link to the Nash equilibrium. Further, our sole aim being that with this mapping the fixed point, i.e., ''Nash equilbrium`` remains unchanged. In this way we get the fixed point of the two player game mapped to the two spin Ising payoff matrix and thus via the ''J`` and ''h`` factors to the magnetization of the infinite spin Ising model. The temperature $T$ in social dilemmas is interpreted as amount of randomness allowed in  selection of strategies by players. $T \to \infty$ implies that players choose their strategies at random while $T \to 0$ implies that no randomness is allowed in selection of strategies by the players. 
One should note that the procedure outlined here is restricted to only mapping the Nash equilibrium of any social dilemma game to the spin-1/2 Ising model. However, using this approach one cannot map the Pareto optimum of the social dilemma game into the spin-1/2 Ising model. The reason being that extra terms (called altruistic terms) have to be added to the Ising Hamiltonian to determine the Pareto optima but as mentioned in Ref.~\cite{Galam2010} itself, this kind of Hamiltonian isn't possible and further there exists no clear interpretation of these extra terms in statistical mechanics (see Ref.~\cite{Galam2010}, sections 4.2 \& 4.3). In Ref.~\cite{Galam2010} it has been concluded that addition of such terms implies the existence of either the individual energies or the Hamiltonian, but not both together, thus we have refrained from including the Pareto optima of the game in our analysis.  

To derive game susceptibilities, we differentiate game magnetization $m_g$ in Eq.~(\ref{genmag}) by the four payoffs to obtain their respective susceptibilities. In general, there will be four susceptibilities corresponding to each of four payoff parameters $a,b,c,d$ of the game as
\begin{eqnarray}
	\label{gensus}
	\chi_a &=& \pdv{m_g}{a}, \; \chi_b = \pdv{m_g}{b}, \; \chi_c = \pdv{m_g}{c}, \; \chi_d = \pdv{m_g}{d}.
\end{eqnarray}
Since $m_g = n_{s_1} - n_{s_2}$, where $n_{s_1}$ and $n_{s_2}$ are the fraction of players selecting strategy $s_1$ and $s_2$ respectively, and $n_{s_1} + n_{s_2} = 1$, we can write the game susceptibility in terms of fraction of players selecting a particular strategy in response to change in payoffs as
\begin{eqnarray}
    \chi_u &=& 2 \pdv{n_{s_1}}{u} \label{gensusprop}
\end{eqnarray}
where, $u$ can be any of the payoff parameters and $n_{s_1}$ is the fraction of players playing  strategy $s_1$. In the following sections, we take a look at thermodynamic susceptibilities in context of Prisoner's dilemma, Hawk-Dove game and finally the Quantum Prisoner's Dilemma(QPD). 
\section{Prisoner's Dilemma}
\subsection{Game magnetization in thermodynamic limit of Prisoner's Dilemma}
Prisoner's dilemma consists of two suspects who have been caught by police and are being separately interrogated for their crimes. Each suspect has two choices, to cooperate (C) with other suspect and not confess to the crime or to defect (D) against other suspect and blame him/her for the crime. 
The payoff matrix is
\begin{eqnarray}
S &=& \left(\begin{array}{c||c|c}
& C & D \\ \hhline{=#=|=}
C & r,r & s,t \\ \hline 
D & t,s & p,p \\
\end{array}\right), \label{pdmatrix}
\end{eqnarray}
where $r$ is reward, $t$ is temptation, $s$ is sucker's payoff and $p$ is punishment with the condition on parameters being $t>r>p>s$. The standard values for $r,s,t,p$ are $r = 3$, $t = 5$, $s = 0$ and $p = 1$ \cite{Poundstone1992}. Of course, as long as inequality $t>r>p>s$ is respected, we can vary the payoffs as $0 \leq s<1$, $1\leq p<3$, $3 \leq r < 5$ and $t \geq 5$. The payoff matrix is understood as follows; payoff of reward $r$ implies a jail time of about 1 year, temptation $t$ implies no jail time for suspect. The sucker's payoff $s$ implies a life term and punishment payoff $p$ implies a jail time of 10 years. It can be easily seen from comparison of general payoff matrix(\ref{gendoublepayoffmatrix}) and payoff matrix of Prisoner's dilemma(\ref{pdmatrix}) that $a = r$, $b = s$, $c = t$ and $d = p$. Each suspect in this case is better off by defecting, since if one player defects while  other cooperates then cooperating player will have a greater loss, hence Nash equilibrium is to defect, irrespective of the other player's choice. But prisoners can definitely do better if both of them choose to cooperate, hence the dilemma.
We get equivalent $J$ and $h$ parameters for the Prisoner's Dilemma from Eq.~(\ref{genrelation}) as
\begin{eqnarray}
J &=& \frac{r-t+p-s}{4} \;\;\;\mbox{ and }  \;\;\; h = \frac{r+s-t-p}{4}. \label{maprel}
\end{eqnarray}
Thus, game magnetization which is effectively the difference between number of cooperators($n_C$) and defectors$n_D$, i.e.,  $m_{g}=n_{C}-n_{D}$ in thermodynamic limit of Prisoner's dilemma is
\begin{eqnarray}
    m_g &=& \frac{\sinh(\frac{r+s-t-p}{4 T})}{\sqrt{\sinh^2(\frac{r+s-t-p}{4 T}) + e^{- \frac{(r-t+p-s)}{T}}}}. \label{gamemag}
\end{eqnarray} In case of Prisoner's dilemma,  the Nash equilibrium at $T\rightarrow 0$ gives $m_{g}\rightarrow -1$(see Eq.~28) implying $n_{C}=0, n_{D}=1$, i.e., all players regardless of whether they are neighbouring or not choose exactly identical strategies, which in this context is Defect.
 At $T \to \infty$, $m_g \to 0$ as players select their strategies at random, leading to equal proportion of cooperators and defectors. 

\subsection{Thermodynamic Susceptibility in Prisoner's Dilemma \label{varpris}}
Following definition of susceptibility for general infinite player game given in \cref{gensus}, we have four susceptibilities associated with each of four payoffs, namely reward susceptibility $\chi_r$, punishment susceptibility $\chi_p$, temptation susceptibility $\chi_t$ and sucker's susceptibility $\chi_s$. The four susceptibilities are directly proportional to the net change in number of cooperators as - 
\begin{equation}
    \chi_a = 2 \pdv{n_C}{a},
\end{equation}
where $a$ is one of the four payoff's and $n_C$ is number of cooperators. In the following sections, we address all four susceptibilities. 
\subsubsection{Reward Susceptibility}
The expression for reward susceptibility $\chi_r$ is
\begin{align}
	\chi_r = \pdv{m_g}{r} = \frac{e^{\frac{s+t}{T}} \left(2 y +\cosh \left(\frac{-p+r+s-t}{4 T}\right)\right)}{4 T \sqrt{e^{\frac{-p-r+s+t}{T}}+ y^2 } \left(e^{\frac{p+r}{T}}  y^2 +e^{\frac{s+t}{T}}\right)}, 
\end{align}
where $y = \sinh \left(\frac{-p+r+s-t}{4 T}\right)$. In limit $T \to 0$, $\chi_r \to 0$. This is so because for $T \to 0$, all players choose to defect while in limit $T \to \infty$, $\chi_r \to 0$ as all players choose their strategies at random. Hence, fraction of players in switching their strategies is completely random and net change is zero on average. Compared to game magnetization ($m_g$) in which $n_{D}> n_{C}$, regardless of the reward($r$). Game susceptibility ($\chi_r$) changes sign, i.e., $\frac{\partial n_{C}}{\partial r} > \frac{\partial n_{D}}{\partial r}$. 

\begin{figure*}
	\centering
	\includegraphics[width= 0.80\linewidth]{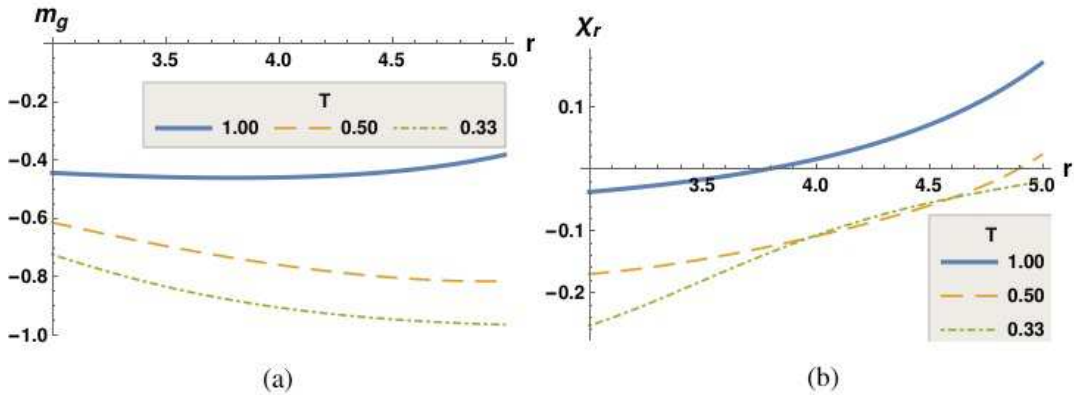}
	\caption{Plots of (a) game magnetization $m_g$, and (b) reward susceptibility $\chi_r$ versus reward payoff ($r$) for different game temperatures $T$. Rest of payoffs are: $t = 5, s = 0$ and $p = 1$. \label{pd_r}}
\end{figure*}

The plot in Fig.~\ref{pd_r}(b) depicts variation of reward susceptibility as a function of reward $r$ in Prisoner's dilemma. The number of  players who switch their strategies in response to change in reward is heavily dependent on game temperature $T$. At lower game temperatures, reward susceptibility is negative while at higher game temperatures reward susceptibility can cross over to positive values, implying that the turnover of players from defect to cooperate exceeds the turnover of players from cooperate to defect. This positive value of susceptibility implies that at higher game temperatures players are vulnerable to increase in reward. A small increase in reward makes the players switch to cooperate. From Fig.~\ref{pd_r}(a), the plot of game magnetization($m_g$), this isn't obvious, since $m_{g} < 0$ always, irrespective of the reward.  The plot of game susceptibility($\chi_r$) show that players are always vulnerable to reward. Any increase in reward makes them change faster to cooperate than to defect. 
\subsubsection{Punishment Susceptibility}
\begin{figure*}[ht]
	\centering
	\includegraphics[width=0.80\linewidth]{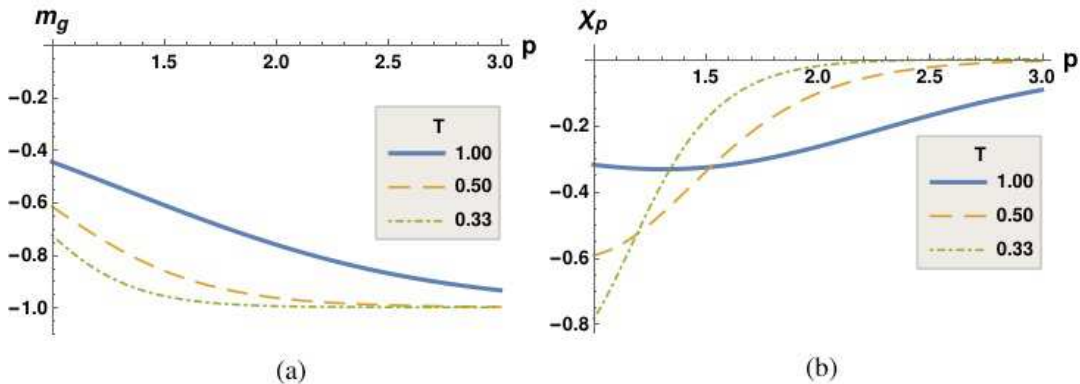}
	\caption{Plots of (a) game magnetization $m_g$ and (b) punishment susceptibility $\chi_p$ versus punishment payoff $p$ for different game temperatures $T$. Rest of payoffs, $t = 5, s = 0$, and $r = 3$ \label{pd_p}}
\end{figure*}

Expression of punishment susceptibility $\chi_p$ is
\begin{align}
	\chi_p = \pdv{m_g}{p} = -\frac{e^{\frac{s+t}{T}} \left(\cosh \left(\frac{-p+r+s-t}{4 T}\right)-2 y \right)}{4 T \sqrt{e^{\frac{-p-r+s+t}{T}}+ y^2} \left(e^{\frac{p+r}{T}}  y^2+e^{\frac{s+t}{T}}\right)}
\end{align}
where $y = \sinh \left(\frac{-p+r+s-t}{4 T}\right)$. In the limit of $T \to 0$, $\chi_p \to 0$ as all players choose to defect. From the two-player case, it is evident that if any player switches to cooperate, they face a loss. On the other hand, in limit of $T \to \infty$, we find that $\chi_p \to 0$, as all players change their strategies at random, which leads to net zero turnover of players, i.e., changing their strategies, on average. The plot in Fig.~\ref{pd_p}(b) depicts variation of punishment susceptibility as a function of punishment $p$ in Prisoner's dilemma. Punishment susceptibility is always negative in response to change in punishment as shown in Fig.~\ref{pd_p}(b), which implies that players prefer to switch to defect. Like the game magnetization($m_g$), punishment susceptibility($\chi_p$) also doesn't change with response to $p$, implying no further information is gained from the susceptibility.  Also, it can be observed that punishment susceptibility diminishes as punishment increases since most players choose defect, and hence the net fraction of players who can switch to defect decreases. Any increase in punishment in PD doesn't lead to any change in player's behavior. 
\subsubsection{Temptation Susceptibility}
\begin{figure*}[ht]
	\centering
	\includegraphics[width= 0.80\linewidth]{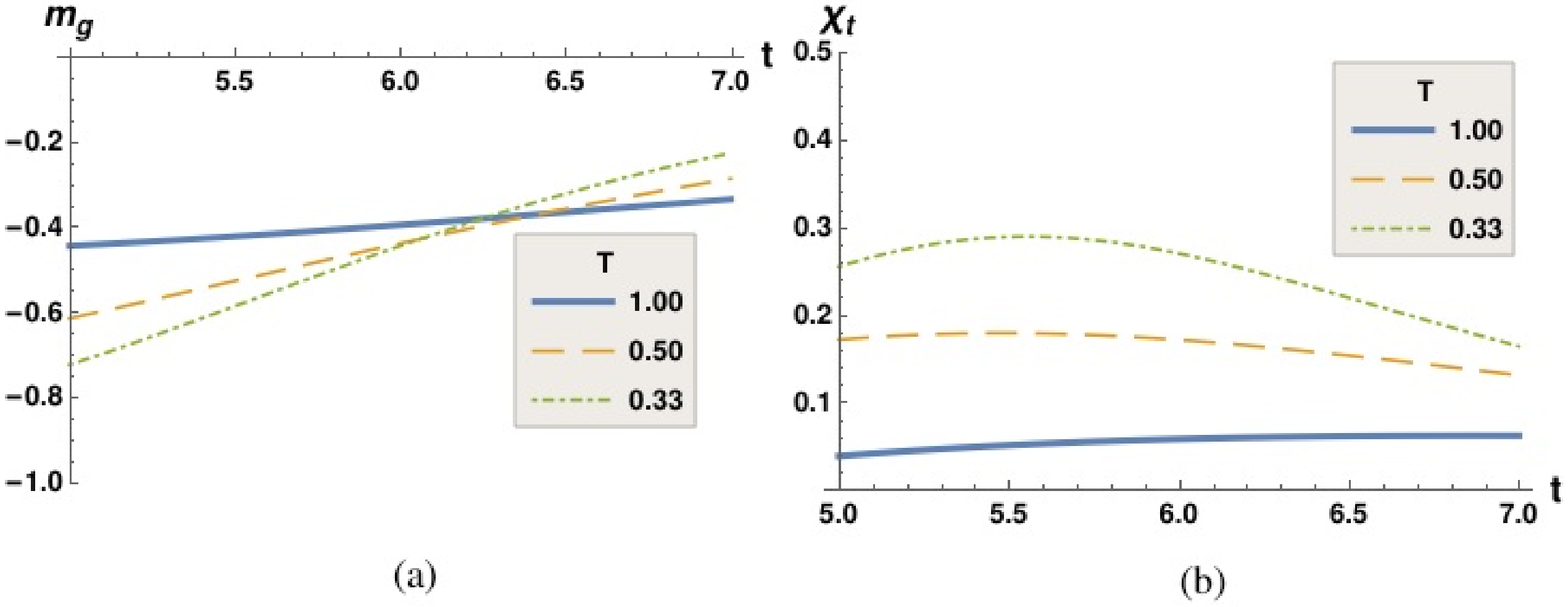}
	\caption{Plots of (a) game magnetization $m_g$ and (b) temptation susceptibility $\chi_t$  versus temptation payoff $t$ for different game temperatures $T$. Rest of payoff's are $p = 1, s = 0$ and $r = 3$. \label{pd_t}}
\end{figure*}

The expression for temptation susceptibility $\chi_t$ is
\begin{align}
	\chi_t =\pdv{m_g}{t}= -\frac{e^{\frac{s+t}{T}} \left(2 y +\cosh \left(\frac{-p+r+s-t}{4 T}\right)\right)}{4 T \sqrt{e^{\frac{-p-r+s+t}{T}}+y^2} \left(e^{\frac{p+r}{T}} y^2+e^{\frac{s+t}{T}}\right)} 
\end{align}
where $y = \sinh \left(\frac{-p+r+s-t}{4 T}\right)$. In limit $T \to 0$, $\chi_t \to 0$. Here, in absence of any randomness in strategy selection, all players select Nash Equilibrium (Defect strategy). Hence, net number of players switching their strategies approaches zero. On the other hand, $\chi_t \to 0$ as $T \to \infty$ as all players choose their strategies at random, so net  number of players switching their strategies is zero. Unlike game magnetization ($m_g$) versus $t$ in Fig.~\ref{pd_t}(a), for which $n_{D} > n_{C}$ always, the game susceptibility has opposite sign, meaning $\frac{\partial n_{C}}{\partial t} > \frac{\partial n_{D}}{\partial t}$ thus the  turnover to cooperative behavior is greater due to change in temptation. Thus at the macroscopic level while defectors dominate regardless of temptation, for small changes in  temptation the turnover of players to cooperation always increases. 
The plot in Fig.~\ref{pd_t}(b) shows variation of temptation susceptibility as a function of temptation $t$. The temptation susceptibility is always positive in response to change in temptation, as  in Fig.~\ref{pd_t}(b). This implies that net fraction of players switching to cooperate strategy is always more than the fraction switching to defect strategy. This positive susceptibility results in overall increase in number of cooperators in the game. 
\subsubsection{Sucker's Susceptibility}
\begin{figure*}[ht]
	\centering
	\includegraphics[width=0.80\linewidth]{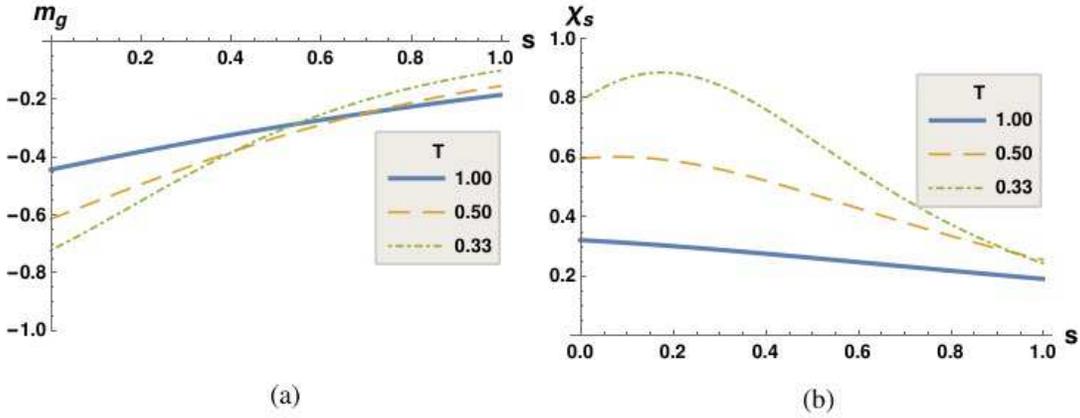}
	\caption{Plots of (a) game magnetization $m_g$ and (b) sucker's susceptibility $\chi_s$ versus sucker's payoff $s$ for different game temperatures $T$. Rest of payoff's are $t=5, r=3$ and $p=1$. \label{pd_s}}
\end{figure*}

The expression for sucker's susceptibility $\chi_s$ is 
\begin{align}
	\chi_s = \pdv{m_g}{s} = \frac{e^{\frac{s+t}{T}} \left(\cosh \left(\frac{-p+r+s-t}{4 T}\right)-2 y \right)}{4 T \sqrt{e^{\frac{-p-r+s+t}{T}}+y^2 } \left(e^{\frac{p+r}{T}} y^2 +e^{\frac{s+t}{T}}\right)} 
\end{align}
where $y = \sinh \left(\frac{-p+r+s-t}{4 T}\right)$. $\chi_s \to 0$ for both limits of $T \to 0$ and $T \to \infty$. In limit of $T \to 0$, we find that players are unwilling to switch their strategies from Nash Equilibrium without any randomness. In limit of $T \to \infty$, players choose their strategies at random, so the net number of players switching their strategies is zero. Again similar to temptation, while at the macroscopic level $n_{D}>n_{C}$ as in Fig.~\ref{pd_s}(b). When looking at microscopic changes to sucker's payoff we see $\frac{\partial n_{C}}{\partial s} > \frac{\partial n_{D}}{\partial s}$ so players are sensitive to change in sucker's payoff, tending to cooperate.
Plot for sucker's susceptibility as a function of sucker's payoff $s$ is given in Fig.~\ref{pd_s}(b). Here, sucker's susceptibility is positive, implying that fraction of players changing from defect to cooperate strategy is higher than the fraction changing from cooperate to defect strategy. This results in net increase in  the number of cooperators.
Finally, in conclusion to this section, we dwell more on ``Temperature''. At zero temperature all players follow the Nash equilibrium strategy which in Prisoner's dilemma is to defect. One can clearly see from Fig.~2(a), as one lowers temperature $T\rightarrow \mbox{low}$, the magnetization $m_{g}\rightarrow -1$ implying all players defect. As temperature increases, more and more players randomly switch from defect to cooperate indicated by the magnetization (net difference between cooperators and defectors) lowering with increasing temperature. Randomness is relative to the ordered state at $T\rightarrow  \mbox{low}$ wherein all players are defectors. The dependence of the game on temperature $T$ is one of the consequences of the mapping between payoffs of the game and the Ising model.
\section{Hawk-Dove Game}
Hawk-Dove Game is another well-known 2-player 2-strategy game\cite{Devos2016} in which  two players contest over a shared resource of value $V$. There is possibility of conflict which inflicts upon both an injury cost  $C$ with, $C>V>0$. Each player can have two possible strategies. The players who adopt "Hawk" strategy show aggressive tendencies and are willing to fight over the resource even at risk of injury. On the other hand, players adopting "Dove" type behavior will not actively try to enter into a conflict, instead, they will try to share the resource or will back away at sign of danger. Denoting Hawk strategy by $H$ and Dove strategy by $D$, the payoff matrix\cite{SMITH1976159} of this game is
\begin{eqnarray}
	\label{ChickenGamePayoff}
	G &=& \left(\begin{array}{c||c|c}
		& Hawk (H) & Dove (D) \\ \hhline{=#=|=}
		Hawk (H) & (\frac{V-C}{2},\frac{V-C}{2}) & (V,0) \\ \hline
		Dove (D) & (0,V) & (\frac{V}{2},\frac{V}{2})
	\end{array} 
	\right).
\end{eqnarray}
Comparing payoff matrix, Eq.~(\ref{ChickenGamePayoff}) with general payoff matrix Eq.~(\ref{gendoublepayoffmatrix}), we have $a=(V-C)/2$, $b=V$, $c=0$ and $d=V/2$. For this game there are two pure strategy Nash Equilibria: $(H,D)$ and $(D,H)$ and a mixed strategy Nash Equilibrium $(\sigma,\sigma)$, where $\sigma = p.H + (1-p).D$, with $p = \frac{V}{C}$ being probability of displaying "Hawk" like behavior. Similar to Prisoner's dilemma, we map "Hawk-Dove Game" to Ising model in the thermodynamic limit. The corresponding transformations are $\lambda = -\frac{V-C}{4}$ and $\mu = \frac{-3V}{4}$ as explained in Section \ref{gengamedef}. The payoff matrix for row player after transformation becomes
\begin{equation}
	\label{Chickengametransform}
	G = \left(\begin{array}{c||c|c}
		& Hawk (H) & turn (D) \\ \hhline{=#=|=}
		Hawk (H) & \frac{V-C}{4} & \frac{V}{4} \\ \hline
		Dove (D) & -\frac{V-C}{4} & -\frac{V}{4}
	\end{array} 
	\right).
\end{equation}
\begin{figure*}[ht]
	\centering
	\includegraphics[width= 0.775\linewidth]{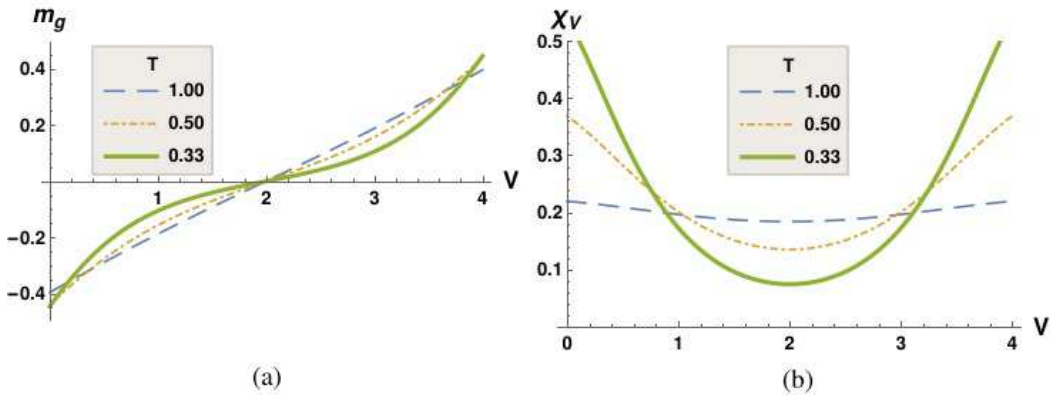}
	\caption{Plots of (a) game magnetization $m_g$ and (b) resource value susceptibility $\chi_V$ versus resource value $V$ for different game temperatures $T$, with cost of injury $C = 4$. \label{hawkv}}
\end{figure*}
\begin{figure*}[ht]
	\includegraphics[width= 0.775\linewidth]{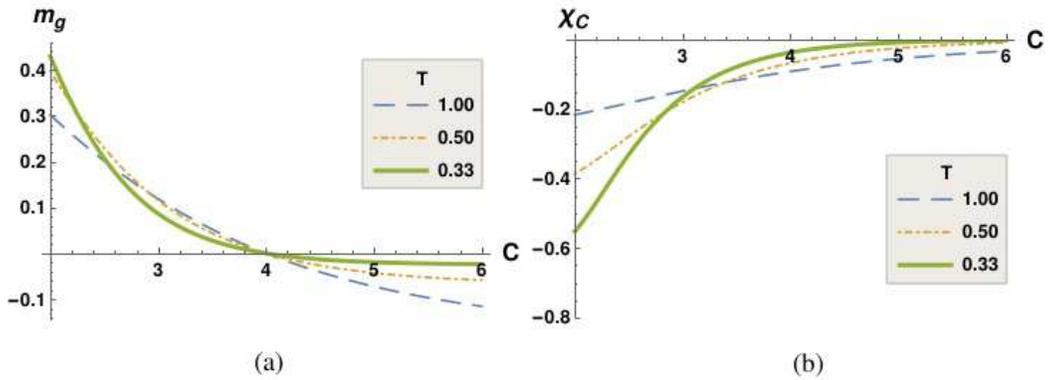}
	\caption{Plots of (a) game magnetization $m_g$ and (b) injury cost susceptibility $\chi_C$ versus injury cost $C$ for different game temperatures $T$, with resource value $V = 2$.\label{hawkc}}
\end{figure*}
The $+1$ spin state is mapped to "Hawk" strategy while $-1$ spin is mapped to "Dove" strategy. From Ising game matrix Eq.~(\ref{energysitematrix}), we have '$J$' and '$h$' factors for the Hawk-Dove game as
\begin{eqnarray}
	J &=& \frac{-C}{8} \;\;\;\; \textnormal{and} \;\;\;\; h = \frac{2V-C}{4}. \label{chickenrel}
\end{eqnarray}
The game magnetization calculated from Eq.~(\ref{genmag}) for Hawk-Dove game is then
\begin{eqnarray}
	m_g&=&\frac{\sinh \left(\frac{2 V-C}{4 T}\right)}{\sqrt{\sinh ^2\left(\frac{2 V-C}{4 T}\right)+e^{\frac{C}{2 T}}}}. \label{hawk_mg}
\end{eqnarray}
At $T \to 0$, $m_g \to 0$, signifying that fraction of Hawks and Doves are equal due to nearest neighbor players selecting opposite strategies to minimize their losses. At $T \to \infty$, $m_g \to 0$ as all players choose their strategies at random, which leads to players selecting Hawk and Dove in equal proportion.
\subsection{Thermodynamic Susceptibility in Hawk-Dove Game}
The game susceptibilities for Hawk-Dove Game are calculated from game magnetization Eq.~(\ref{hawk_mg}) as
\begin{eqnarray}
	\label{hawksus}
	\chi_V &=& \pdv{m_g}{V} \;\;\;\;\textnormal{and}\;\;\;\; \chi_C = \pdv{m_g}{C},
\end{eqnarray}
where $\chi_V$ is resource susceptibility and $\chi_C$ is cost susceptibility.
Further, using $m_g = n_H - n_D$, where $n_H$ and $n_D$ are fraction of players selecting Hawk and Dove strategies respectively, and as $n_H + n_D = 1$, we express the game susceptibilities as
\begin{eqnarray}
	\label{hawksusalt}
	\chi_q &=& 2 \pdv{n_H}{q} \textnormal{   with   } q = C,V .
\end{eqnarray}
\subsubsection{Resource Susceptibility}
The resource susceptibility calculated from Eq.~(\ref{hawksus}) is 
\begin{eqnarray}
	\chi_V &=& \pdv{m_g}{V} = \frac{e^{\frac{C}{2 T}} \cosh \left(\frac{C-2 V}{4 T}\right)}{2 T \left(\sinh ^2\left(\frac{2 V-C}{4 T}\right)+e^{\frac{C}{2 T}}\right)^{3/2}}. \label{chiv}
\end{eqnarray}
In the limit $T \to 0, \chi_V \sim \exp(-T^3)$ for both $V \to C$ and $V \to 0$. 
At $T = 0, \chi_V \propto V$ for $V \to 0$ and $V \to C$. At infinite temperature ($T \to \infty$) $\chi_V \to 0$ due to complete randomness in strategy selection by players. The plot for resource susceptibility $\chi_V$ versus resource value $V$ in Fig.~\ref{hawkv}(b) shows that net the number transitioning their strategies has a minima at $V=C/2$ for all $T$. The susceptibility is at a minimum for $V=C/2$ as it represents an inflexion point. The resource susceptibility $\chi_V$ is always positive, implying that the net turnover of players from Dove to Hawk is higher than the net turnover from Hawk to Dove. Overall, this implies that fraction of players playing Hawk increases as resource value increases but the fraction changing their strategies from dove to hawk like behavior or vice-versa crucially depends on the value of the resource.
\subsubsection{Cost Susceptibility}
The cost susceptibility calculated from Eq.~(\ref{hawksus}) is 
\begin{align}
	\chi_C =\pdv{m_g}{C}= -\frac{e^{\frac{C}{4 T}} \left(\sinh \left(\frac{V}{2 T}\right)+\cosh \left(\frac{V}{2 T}\right)\right)}{4 T \left(\sinh ^2\left(\frac{2 V-C}{4 T}\right)+e^{\frac{C}{2 T}}\right)^{3/2}}.
\end{align}
In the limit $T \to 0$, $\chi_C\sim\exp(-T^2)$ as $V \to C$ and $\chi_C \to 0$ as $V \to 0$. At $T = 0, \chi_C \propto (V-C)$ for $V \to C$, and for $T \to \infty, \chi_C \to 0$ due to complete randomness in strategy selection by players, which results in equal number of Hawks and Doves in the game. The plot of cost susceptibility $\chi_C$ as a function of injury cost $C$ in Fig.~\ref{hawkc}(b) indicates that $\chi_C$ is negative, implying that the net turnover from Hawk to Dove is greater than the turnover from Dove to Hawk. This means that proportion of Dove players increases with increasing cost. Next, we check for a quantum game.
\section{Quantum Prisoner's Dilemma}
The Quantum Prisoner's Dilemma (QPD) game\cite{Eisert1999,Flitney2002,bordg} is the quantum version of Prisoner's dilemma wherein players are each assigned a qubit, which is in a superposition of states $\ket{C}$ and $\ket{D}$, represented in 2D Hilbert space as
\begin{equation}
    \ket{C} = \begin{bmatrix}
        1 \\ 0\\ 
        \end{bmatrix} \;\;\;\;;\;\;\;\; \ket{D} = \begin{bmatrix}
        0 \\ 1\\ 
        \end{bmatrix}.        
\end{equation}
$\ket{C}$ and $\ket{D}$ represent states of cooperation and defection respectively in analogy with the classical PD game. 
The strategy to be employed by each player is given by unitary operator $U(\theta,\phi)$, as
\begin{eqnarray}
    \label{genoperator}
    U(\theta,\phi) &=&  \begin{bmatrix}
        e^{i \phi}\cos(\theta /2) & \sin(\theta /2) \\
        - \sin(\theta /2) & e^{-i \phi}\cos(\theta /2) \\
        \end{bmatrix}.
\end{eqnarray}
Here, $\theta \in [0,\pi]$ and $\phi \in [0, \pi/2]$. 
The classical operations of cooperate and defect can be represented as $U(0,0) = I$ and $U(0,\pi) = X$ respectively ($I$ is identity matrix while $X = \sigma_x$ is the Pauli matrix). 
Before players are allowed to operate their strategies on their respective qubits, their respective qubits are entangled by entanglement operator $\mathcal{J}(\gamma)$. It is given by
\begin{eqnarray}
    \label{J_gamma}
    \mathcal{J}(\gamma) &=& \cos\left(\frac{\gamma}{2}\right) I \otimes I + i \sin\left(\frac{\gamma}{2}\right) Y \otimes Y.
\end{eqnarray} 
where $Y = i \sigma_y$. The game begins by letting both players (say A and B) have a qubit of their own, in $\ket{C}$ state. Next, the entangling operator $\mathcal{J}(\gamma)$ acts upon both qubits to entangle them. After this, both players choose a operator $U$ and apply it to their qubit. Finally, before the qubits are measured, a disentangling operator $J^\dagger(\gamma) $ is applied to the entangled state and payoffs are calculated via taking inner product of states $\ket{CC}, \ket{DC},\ket{CD}$ and $\ket{DD}$ with the final state $\ket{\psi_f}$, and multiplying with corresponding classical payoffs for states, we have payoffs for players A ($\$_A$) and B ($\$_B$) as
\begin{align}
    \$_A &= r P_{CC} + p P_{DD} + t P_{DC} + s P_{CD},\\
 \mbox{and, }   \$_B &= r P_{CC} + p P_{DD} + s P_{DC} + t P_{CD},
\end{align}
where, $P_{CC} = |\braket{CC}{\psi_f}|^2$, $P_{CD} = |\braket{CD}{\psi_f}|^2$, $P_{DC} = |\braket{DC}{\psi_f}|^2$, and $P_{DD} = |\braket{DD}{\psi_f}|^2$, $\ket{\psi_f}$ being final state of the qubits after execution of game. One can also have a quantum operator $Q = iZ$, where $Z = \sigma_z$.
The strategy Q or the quantum strategy is offered to the players in the quantum Prisoner's dilemma (QPD) as an alternative to the classical strategies of cooperate or defect. In terms of Pauli matrices: cooperate(C) is Identity while defect(D) is $\sigma_x$ and quantum is $i\sigma_z=e^{i\frac{\pi}{2}}\sigma_z$. Thus while the quantum superposition state $|+\rangle=\frac{1}{\sqrt{2}}(|0\rangle +|1\rangle)$ is invariant under classical strategies like cooperate $I$ or defect $\sigma_x$, action of quantum strategy Q on $|+\rangle$ gives the orthogonal $|-\rangle=\frac{1}{\sqrt{2}}(|0\rangle - |1\rangle)$ state multiplied by a global phase of $\pi/2$.

\begin{figure}[htbp]
	\centering
	\includegraphics[width = \linewidth]{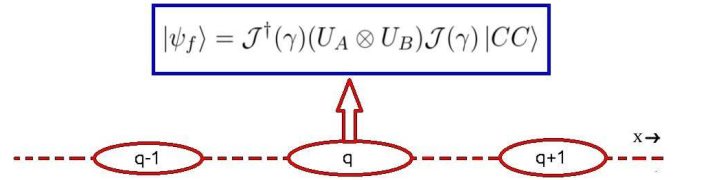} 
	\caption{Extending the Quantum Prisoner's Dilemma to thermodynamic limit. Each site (ellipse) consists of two players who play the two-player quantum prisoner's dilemma. The sites are connected via coupling $J$ and all the sites are influenced by the external field $h$.}
	\label{implement}
\end{figure}

Thus, for QPD game including both classical and quantum strategies, the payoff matrix is 
\begin{eqnarray}
	G &=&  \left(\begin{array}{c||c|c|c}
		&  C & D & Q \\ \hhline{=#=|=|=}
        C & r,r & s,t & l_1,l_1 \\ \hline
        D & t,s & p,p & l_3,l_2  \\ \hline
        Q & l_1,l_1 & l_2,l_3 & r,r \\
    \end{array}\right), \label{quantumpayoff}
\end{eqnarray}
where $l_1 = r \cos^2(\gamma) + p \sin^2(\gamma)$, $l_2 = s \cos^2 (\gamma) + t \sin^2 (\gamma)$ and $l_3 = t \cos^2 (\gamma) + s \sin^2 (\gamma)$. When entanglement between players is maximal ($\gamma = \pi/2$), the payoff matrix with $r = 3, s = 0, p= 1$ and $t = 5$ becomes
\begin{equation}
	G =  \left(\begin{array}{c||c|c|c}
		&  C & D & Q \\ \hhline{=#=|=|=}
        C & 3,3 & 0,5 & 1,1 \\ \hline
        D & 5,0 & 1,1 & 0,5  \\ \hline
        Q & 1,1 & 5,0 & 3,3 \\
	\end{array}\right), 
\end{equation}
and Nash equilibrium is the quantum strategy (Q,Q) with a payoff of (3,3) for each player. 
For extending the game to thermodynamic limit by mapping to Ising model, we must perform some modifications to game setup to incorporate entanglement between players. Each site in the Ising chain is occupied by two players who play a two-player QPD. Each site interacts with its nearest-neighbor site via classical coupling $J$. The sites in Ising chain are influenced by equivalent external factor $h$ to behave similarly. A schematic to understand the setup of QPD in thermodynamic limit is given in Fig.~\ref{implement}.
 Now, QPD game is a 3-strategy game. However, there exists no meaningful and consistent method to map a 3-strategy social dilemma game to an analytically solvable spin-1 Ising model to , i.e., an Ising model for spin-1 states (0,1,-1).. Hence, we decided to split the QPD into two sub-problems to better understand the advantages of the quantum strategy against the classical strategies of cooperate and defect. Further, since magnetization in Ising model compares the fraction of up spins with the number of down spins in the chain, we break the payoff matrix of the QPD(\ref{quantumpayoff}) into $ 2 \times 2$  matrix such that the quantum strategy can be compared against a classical strategy.

So, we have two cases for the QPD, Quantum vs. Cooperate and Quantum vs. Defect case\cite{Sarkar2018a}. 
The case of Quantum vs. Cooperate has payoff matrix given as - 
\begin{eqnarray}
G &=& \left(\begin{array}{c||c|c}
& Q & C \\ \hhline{=#=|=}
Q & r & r \cos^2(\gamma) + p \sin^2(\gamma) \\
C & r \cos^2(\gamma) + p \sin^2(\gamma) & r \nonumber\\
\end{array}\right).\\ \label{QvsC}
\end{eqnarray}
Using a method similar to derivation of Eq.~(\ref{genrelation}), we obtain relations for the '$J$' and '$h$' as 
\begin{equation}
	J = \frac{(r-p)\sin^2 (\gamma)}{2} \;\;\;\; \mbox{ and } \;\;\; h = 0.
\end{equation}
 It follows from here that corresponding susceptibilities are zero for Quantum vs. Cooperate. Hence, we do not further analyze this case and concentrate on the Quantum vs. Defect case in the thermodynamic limit. 
For case of Quantum vs. Defect case, corresponding reduced payoff matrix is given as
\begin{eqnarray}
G = \left(\begin{array}{c||c|c}
& Q & D \\ \hhline{=#=|=}
Q & r & s \cos^2(\gamma) + t \sin^2(\gamma) \\ \hline
D & t \cos^2(\gamma) + s \sin^2(\gamma) & p \\
\end{array}\right). \label{QvsD}
\end{eqnarray}
We derive the '$J$', '$h$' relations by using a similar procedure as used in \cref{genrelation} as
\begin{eqnarray}
	J &=& \frac{r+p-t-s}{4} \;\;\;\;\;\mbox{ and }\;\;\;\;\; h = \frac{r-p + (s-t)\cos(2\gamma)}{4}. \label{qpdrel}
\end{eqnarray}
The game magnetization for the Quantum vs. Defect is then
\begin{eqnarray}
	m_g &=& \frac{\sinh \left(\frac{-p+r+\cos (2 \gamma ) (s-t)}{4 T}\right)}{\sqrt{\sinh^2  \left(\frac{-p+r+\cos (2 \gamma ) (s-t)}{4 T}\right) +e^{\frac{-p-r+s+t}{T}}}} . \label{gamma_mag}
\end{eqnarray}
\begin{figure}[htbp]
    \centering
	\includegraphics[width = 0.80\linewidth]{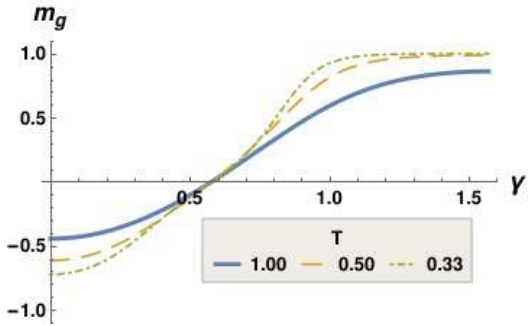}
    \caption{Variation of magnetization as a function of entanglement $\gamma$. Here, $r=3,s=0,p=1,t=5$}
    \label{mg_gamma}
\end{figure}
From expression of game magnetization, we find that magnetization switches from negative to positive as entanglement increases.   
The transition from the "quantum``  to "defect'' occurs at 
\begin{eqnarray}
	\gamma_0 &=& \dfrac{1}{2} \arccos \dfrac{r-p}{t-s}. \label{gamma0}
\end{eqnarray} 
Now, $\gamma_0 \approx 0.579$ for $r=3,s=0,p=1$ and $t=5$ as shown in(\ref{mg_gamma}) and is independent of $T$. It should be however, noted that $\gamma_0$ does not mark a point of phase transition as neither the magnetization is discontinuous there, nor is it accompanied by any form of divergent susceptibility, as will be seen in \cref{QPDsus}. For the range of payoffs of $0 \leq s < 1, \; 1 \leq p < 3, \; 3 \leq r < 5, \; 5 \leq t < 7$, the point of  transition $\gamma_0$ lies in the range $(0,\pi/4)$.

The game susceptibilities for QPD are derived by taking derivative of game magnetization $m_g$ with each of four payoff parameters. The susceptibilities for QPD are proportional to the net turnover  fraction of quantum players, i.e., $\chi_k = 2 \pdv{n_Q}{k}$ where $k$ can be either $r,s,t,p$ and $n_C$ is the fraction of cooperators. We do not define any susceptibility for entanglement parameter $\gamma$ as entanglement among the qubits is not controlled by players' actions but is introduced via the game setup. Further, when entanglement $\gamma$ is zero, the game becomes equivalent to the classical Prisoner's Dilemma.
\section{Susceptibility in thermodynamic limit of Quantum Prisoner's Dilemma \label{QPDsus}}
We analyze the four susceptibilities as a function of four payoffs as well as entanglement in this section. In this game, the four susceptibilities are defined in same manner as susceptibilities defined in \ref{varpris}. We analyze the susceptibilities at finite temperature in \crefrange{reward}{pun}. The special case of $T \to 0$ and $T \to \infty$ case are analyzed in\ref{T0}.
\subsection{Reward Susceptibility \label{reward}}
\begin{figure*}[htbp]
    \includegraphics[width = \linewidth,keepaspectratio]{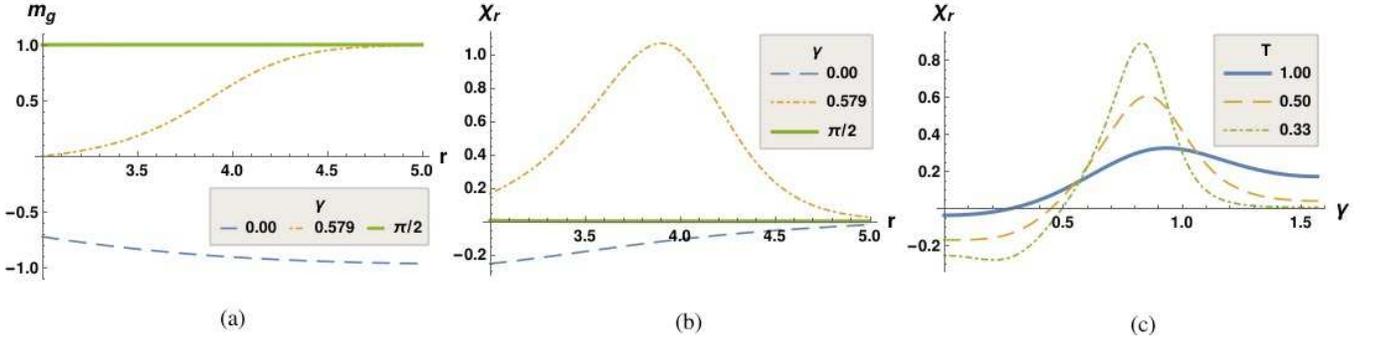}
    \caption{ Plot of game magnetization (a) and reward susceptibility (b) as a function of reward $r$ for different values of entanglement $\gamma$ at game temperature $T=0.33$ (payoffs: $T=0.33, s=0,t=5 \; \& \; p=1$).(c) Plot of reward susceptibility $\chi_r$ versus entanglement $\gamma$ (payoffs: $r=3,s=0,t=5 \; \& \; p=1$).}
    \label{chi_r}
\end{figure*}

The reward susceptibility for QPD calculated from\ref{gamma_mag} is
\begin{eqnarray}
	\chi_r &= \pdv{m_g}{r} = \frac{e^{\frac{s+t}{T}} \left(2 y + \cosh \left(\frac{-p+r+\cos (2 \gamma ) (s-t)}{4 T}\right)\right)}{4 T \sqrt{y^2 +e^{\frac{-p-r+s+t}{T}}} \left(e^{\frac{p+r}{T}} y^2 +e^{\frac{s+t}{T}}\right)} \label{qpdrewsus}
\end{eqnarray}
where $y = \sinh \left(\frac{-p+r+\cos (2 \gamma) (s-t)}{4 T}\right)$. 
In Fig.~\ref{chi_r}(a), we plot the variation of game magnetization versus reward $r$ at $ T = 0.33$ for different $\gamma$. We observe that majority of players for $\gamma_{0} < \gamma < \pi/2$ choose quantum over defect strategy.  The plots of reward susceptibility as a function of reward $r$ is given in Fig.~\ref{chi_r}(b) for particular values of entanglement at $T = 0.33$ while variation of reward susceptibility versus entanglement $\gamma$ for specific payoffs is given in Fig.~\ref{chi_r}(c) at different game temperatures. 
From Fig.~\ref{chi_r}(b), we observe that increasing reward for higher entanglement values increases the net fraction of players who choose quantum strategy as reward susceptibility is positive, implying that the net turnover of players from defect to quantum is higher than the net turnover from quantum to defect. This implies that fraction of players selecting a quantum strategy increases, as can be seen in Fig.~\ref{chi_r}(a). At maximal entanglement, reward susceptibility is zero at low $T$ as all players choose quantum strategy as shown in Fig.~\ref{chi_r}(b). In Fig.~\ref{chi_r}(c), for $\gamma < \gamma_0$, reward susceptibility is negative implying that for low values of entanglement, net fraction of players changing to defect is more while for $\gamma > \gamma_0$, the reward susceptibility is positive, implying that reward promotes  switching to quantum at higher entanglement. At maximal entanglement, $\chi_r \to 0$ for low $T$, implying that the effect of reward on the players becomes negligible. Further, there is no phase transition as the susceptibility does not diverge at any point.
\begin{figure*}[htbp]
    \includegraphics[width = \linewidth,keepaspectratio]{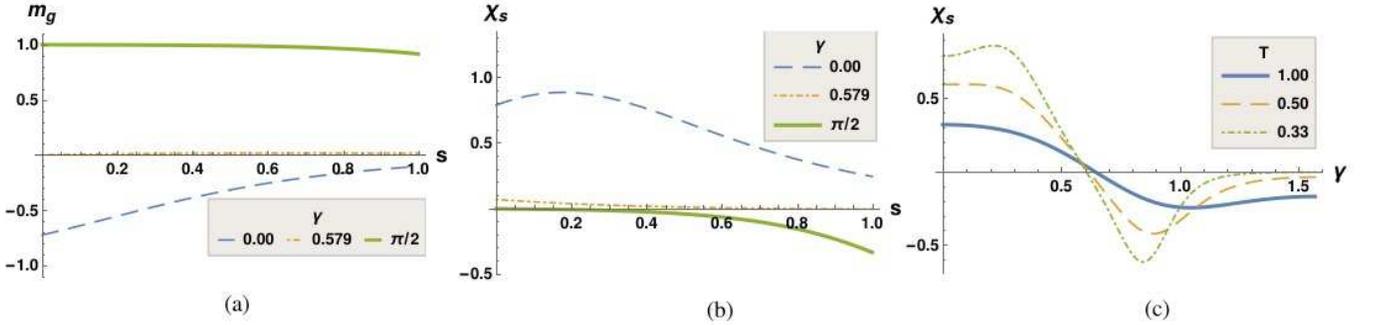}
    \caption{Plot of game magnetization(a) and sucker's susceptibility (b) as a function of sucker's payoff $s$ for different values of entanglement $\gamma$ at $T=0.33$, (payoffs: $r=3, t=5 \; \& \; p=1$). (c) Plot of sucker's susceptibility $\chi_s$ versus entanglement $\gamma$ (payoffs: $r=3,s=0,t=5 \; \& \; p=1$).}
    \label{chi_s}
\end{figure*}
\begin{figure*}[htbp]
    \includegraphics[width = \linewidth,keepaspectratio]{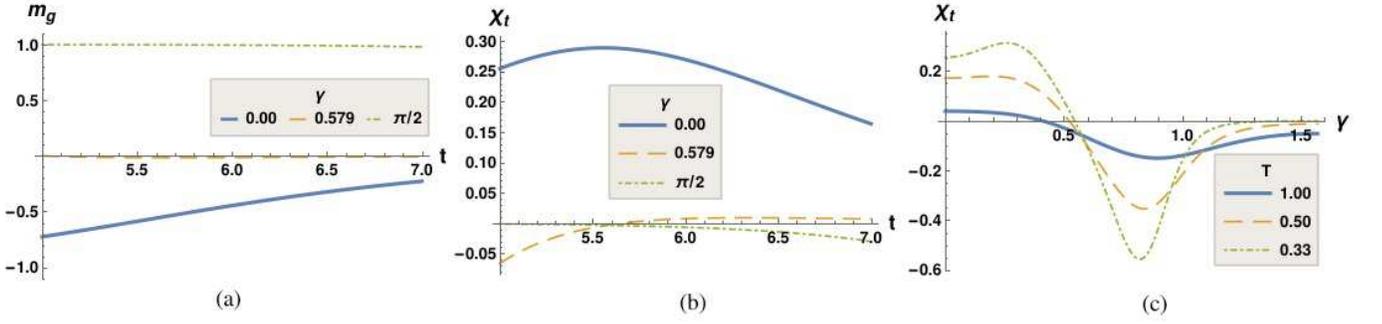}  
    \caption{Plot of game magnetization(a) and temptation susceptibility(b) as a function of temptation $t$ for different values of entanglement $\gamma$ at $T=0.33$ (payoffs: $ r=3, s=0 \; \& \; p=1$).(c) Plot of temptation susceptibility $\chi_t$ versus entanglement $\gamma$  (payoffs: $r=3,s=0,t=5 \; \& \; p=1$).}
    \label{chi_t}
\end{figure*}
\begin{figure*}[htbp]
    \includegraphics[width = \linewidth,keepaspectratio]{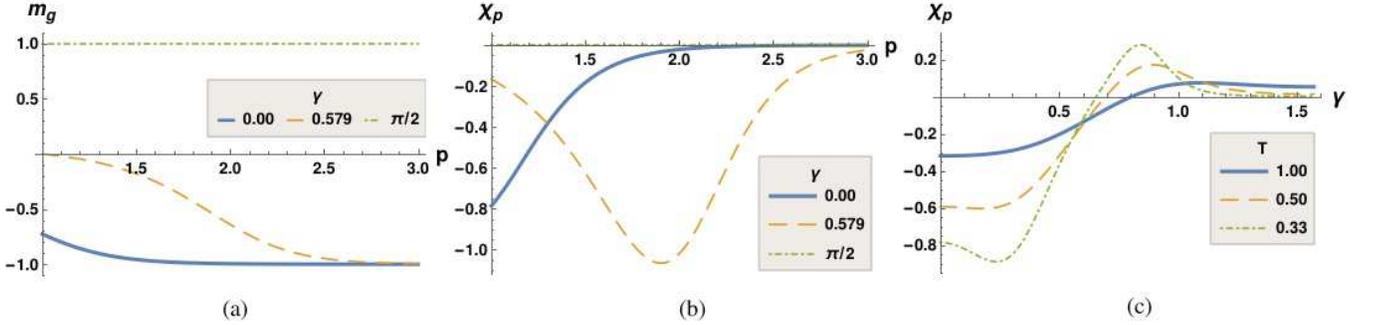}
    \caption{Plot of game magnetization(a) and punishment susceptibility (b) as a function of punishment $p$ for different values of entanglement $\gamma$ at $T=0.33$ (payoffs:  $r=3, s=0 \; \& \; t=5$).(c) Plot of punishment susceptibility $\chi_p$ versus entanglement $\gamma$ (payoffs:  $r=3,s=0,t=5 \; \& \; p=1$).}
    \label{chi_p}
\end{figure*}
\subsection{Sucker's Susceptibility \label{suck}}
The sucker's susceptibility for QPD calculated from \cref{gamma_mag} is -
\begin{eqnarray}
	\chi_s &=& \pdv{m_g}{s} = \frac{e^{\frac{s+t}{T}} \left(\cos (2 \gamma ) \cosh \left(\frac{-p+r+\cos (2 \gamma ) (s-t)}{4 T}\right)-2 y \right)}{4 T \sqrt{y^2 +e^{\frac{-p-r+s+t}{T}}} \left(e^{\frac{p+r}{T}} y^2 +e^{\frac{s+t}{T}}\right)} \label{qpdsucksus}
\end{eqnarray}
where $y = \sinh \left(\frac{-p+r+\cos (2 \gamma ) (s-t)}{4 T}\right)$. 
In Fig.~\ref{chi_s}(a), we observe that at $\gamma = \gamma_0$, the fraction of quantum and defect population in the game are weakly affected by change in sucker's payoff. On the other hand, at maximal entanglement ($\gamma = \pi/2$), we find that game magnetization decreases with increase in sucker's payoff $s$, implying that sucker's payoff increases the fraction of defectors in game.  
The plots of sucker's susceptibility as a function of sucker's payoff $s$ is given in Fig.~\ref{chi_s}(b) for some values of entanglement while variation of sucker's susceptibility with entanglement $\gamma$ is given in Fig.~\ref{chi_s}(c) for different game temperatures. At $\gamma_0$ for low $T$, sucker's susceptibility is almost zero, as can be seen in Fig.~\ref{chi_s}(b). This is so as fraction of quantum and defect players are almost equal and fixed. At maximum entanglement, sucker's susceptibility becomes negative and decreases as sucker's payoff is increased. This implies that the net  turnover from quantum to defect is higher than vice-versa, which increases proportion of defectors. In Fig.~\ref{chi_s}(c), for $\gamma < \gamma_0$, sucker's susceptibility is positive, implying that at low entanglement, change in sucker's payoff promotes players to select quantum, while for $\gamma > \gamma_0$, sucker's susceptibility transitions to negative values, implying that sucker's payoff at higher entanglement, makes players select defect over quantum strategy. At maximal entanglement, the effect of sucker's payoff on players is non-existent at low $T$. Further, there is no phase transition in the game at finite $T$ as susceptibility does not diverge at any value of entanglement.  
\subsection{Temptation Susceptibility \label{tempt}}
The temptation susceptibility for QPD calculated from \cref{gamma_mag} is
\begin{eqnarray}
	\chi_t &=& \pdv{m_g}{t} = -\frac{e^{\frac{s+t}{T}} \left(2 y +\cos (2 \gamma ) \cosh \left(\frac{-p+r+\cos (2 \gamma ) (s-t)}{4 T}\right)\right)}{4 T \sqrt{y^2 +e^{\frac{-p-r+s+t}{T}}} \left(e^{\frac{p+r}{T}} y^2 +e^{\frac{s+t}{T}}\right)}, \label{qpdtemptsus}
\end{eqnarray}
where $y = \sinh \left(\frac{-p+r+\cos (2\gamma) (s-t)}{4 T}\right)$.
The game magnetization as a function of temptation $t$ at game temperature $T = 0.33$ for different entanglement values is given in Fig.~\ref{chi_t}(c). We observe that at $\gamma = \gamma_0$, the game magnetization($m_g$) is nearly constant in response to change in temptation $t$ while at $\gamma = \pi/2$, i.e., maximal entanglement, $m_g$ decreases slightly as temptation increases, implying that temptation aids the defector population. The plots of temptation susceptibility as a function of temptation $t$ is given in Fig.~\ref{chi_t}(b) for some entanglement values while variation of temptation susceptibility with entanglement $\gamma$ is shown in Fig.~\ref{chi_t}(c) for particular game temperatures $T$.
In Fig.~\ref{chi_t}(b), we observe that temptation susceptibility at entanglement $\gamma_0$ is weakly dependent on temptation $t$. At maximal entanglement, however, temptation susceptibility is negative and decreases as temptation increases, implying that fraction of defectors increases as temptation is increased.
In Fig.~\ref{chi_t}(c), for $\gamma < \gamma_0 $, players prefer to switch to quantum as temptation susceptibility is positive, while for $\gamma > \gamma_0$, players switch to defect preferentially as temptation susceptibility becomes zero at maximal entanglement, implying players are not influenced by temptation. For low $T$, temptation susceptibility becomes zero at maximal entanglement, implying that players are not influenced by temptation. There is no phase transition involved for finite $T$ since susceptibility does not diverge for any value of $\gamma$.  
\subsection{Punishment Susceptibility \label{pun}} 
\begin{figure*}[htbp]
	\centering
	\includegraphics[width = \linewidth]{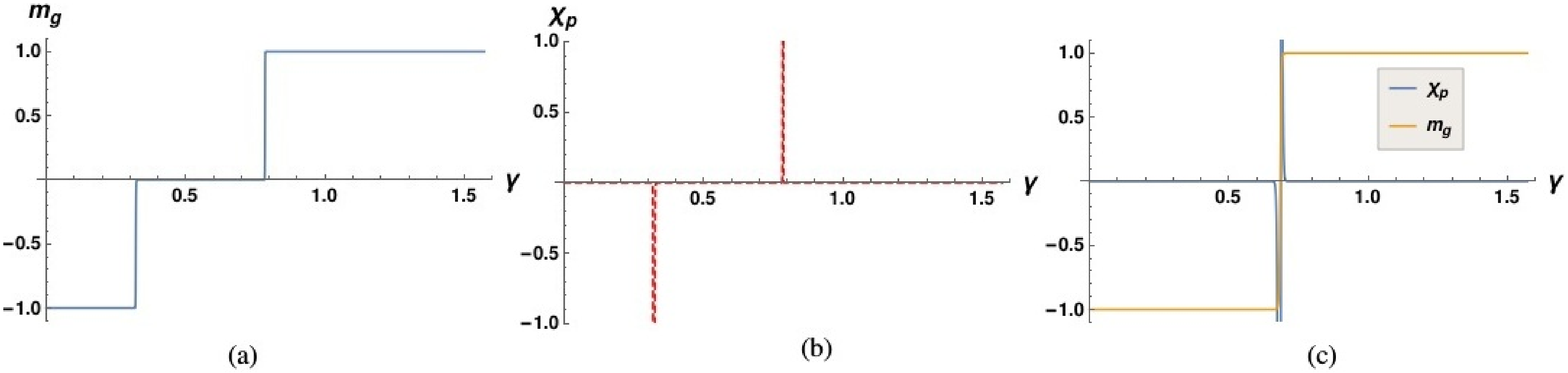}
	\caption{Plot of (a) game magnetization and (b) punishment susceptibility versus entanglement $\gamma$ in the $T \to 0$ limit. (All the other susceptibilities (reward, sucker's and temptation) show identical behavior to punishment susceptibility.) Here, $r =3, s=0,t=5$ and $p=1$. (c) plots the punishment susceptibility and the game magnetization for a case where the random phase does not exist. Here, $r =3, s=0,t=5$ and $p=2$.}
	\label{zerotemp}
\end{figure*}
The punishment susceptibility for QPD calculated from Eq.~(\ref{gamma_mag}) is 
\begin{eqnarray}
	\chi_p &=& \pdv{m_g}{p} = -\frac{e^{\frac{s+t}{T}} \left(\cosh \left(\frac{-p+r+\cos (2 \gamma ) (s-t)}{4 T}\right)-2 y \right)}{4 T \sqrt{y^2 +e^{\frac{-p-r+s+t}{T}}} \left(e^{\frac{p+r}{T}} y^2 +e^{\frac{s+t}{T}}\right)} \label{qpdpunsus}
\end{eqnarray}
where $y = \sinh \left(\frac{-p+r+\cos (2 \gamma ) (s-t)}{4 T}\right)$.
The game magnetization $m_g$ versus the punishment $p$ for $T = 0.33$ and at different entanglement values is shown in Fig.~\ref{chi_p}(a). We observe that at $\gamma = \gamma_0$, punishment promotes the population of defectors in the game. At maximal entanglement, punishment has no effect on the players in the game as game magnetization is constant.
Plots for punishment susceptibility as function of punishment $p$ are given in Fig.~\ref{chi_p}(b) for some values of entanglement while variation of punishment susceptibility with entanglement $\gamma$ is given in Fig.~\ref{chi_p}(c) for some game temperatures.
We observe from Fig.~\ref{chi_p}(b) that punishment susceptibility is negative at entanglement $\gamma_0$, implying that net turnover of players from quantum to defect is more than vice-versa. At maximal entanglement, punishment does not affect players as punishment susceptibility is zero. In Fig.~\ref{chi_p}(c), punishment susceptibility is positive for $\gamma >\gamma_0$, implying punishment $p$ promotes transition to quantum strategy. On the other hand, in the domain $\gamma < \gamma_0$, punishment susceptibility is negative, implying that at low entanglement, punishment induces players to defect. The finiteness of susceptibility for all entanglement $\gamma$ implies the absence of a phase transition in the game. In the next section, we look at the susceptibilities and the magnetization in the zero temperature limit. 
\subsection{Payoff susceptibilities in QPD for $T \to 0$ and $T \to \infty$ limit. \label{T0}} 
In the previous four subsections, we have analyzed the payoff susceptibilities in case of QPD for a wide range of parameters. We have explicitly looked at both low and high $T$ in case of Fig.~\hyperref[chi_r]{10}(c), \hyperref[chi_s]{11}(c), \hyperref[chi_t]{12}(c) and \hyperref[chi_p]{13}(c).
In the limit $T \to \infty$, we find that all four susceptibilities reduce to zero. This is because the players select their strategies at random which leads to net zero turnover of players switching their strategies. 

  Plotting the game magnetization in Fig.~\ref{zerotemp}(a) in the limit $T \to 0$, we observe that there are two transition points, where the magnetization of the game changes. At entanglement $\gamma_1$, we find that the magnetization changes from $-1$ to $0$, while at $\gamma_2$, we observe the magnetization change from $0$ to $+1$, with $\gamma_1 < \gamma_2$. Accordingly, we have three phases in the game, namely, the classical phase which exists for $0 < \gamma < \gamma_1$ in which the classical coupling dominates the choice of players and influences players to choose defect. The second phase exists for the entanglement range of $\gamma_1 < \gamma < \gamma_2$. In this phase, the influence of classical coupling between sites and the entanglement at a site completely cancel each other out, which leads to players selecting their strategies at random. The third phase exists in the range $\gamma_2 < \gamma < \pi/2$ in which the strategy selection in governed by the entanglement, which influences players to select quantum strategy. 
Since the game magnetization curve is discontinuous at $\gamma_1$ and $\gamma_2$ for $T \to 0$, we find that all four susceptibilities diverge for these two points identically. The phase transition is second-order as the susceptibility in the game diverges at $\gamma_1$ and $\gamma_2$ \cite{Baxter1982,Jaeger1998}. The phase transition is similar to the ferromagnet-paramagnet phase transition and the superconductor-normal metal phase transition as the magnetic susceptibility diverges near the critical points, similar to divergence of game susceptibilities near the critical entanglement\cite{Martien}.  

To find an analytic expression for $\gamma_1$ and $\gamma_2$, we solve the equation 
\begin{eqnarray}
	\sinh^2 \left(\dfrac{r-p + \cos(2 \gamma)(s-t)}{4T}\right) &=& \exp(\dfrac{s+t-r-p}{T}) \label{gamma_1_2_condition}
\end{eqnarray}
for $\gamma$ while ensuring that the condition $t>r>p>s$ holds. Eq.~(\ref{gamma_1_2_condition})  can be obtained from either magnetization(\ref{gamma_mag}) or the susceptibility expressions (\ref{qpdrewsus},\ref{qpdsucksus},\ref{qpdtemptsus}) and (\ref{qpdpunsus}). The analytic forms for $\gamma_1$ and $\gamma_2$ are 
\begin{equation}
	\gamma_1 = \dfrac{1}{2} \arccos\left(\dfrac{3p+r-2(s+t)}{s-t}\right), \label{gamma1}\end{equation}
\begin{equation}
	\mbox { and }\gamma_2 =\dfrac{1}{2} \arccos\left(\dfrac{3r-p+2(s-t)}{s-t}\right). \label{gamma2}
\end{equation}
with $\gamma_1 \approx 0.321$ and $\gamma_2 \approx 0.785$ in Figs.~\ref{zerotemp}(a) and \ref{zerotemp}(b) respectively. The solutions of Eq.~(\ref{gamma_1_2_condition}) are valid as long as the condition $p+r<s+t$ is satisfied. This provides the condition for the existence of random phase in the game. If the condition $p+r<s+t$ is not satisfied, then there are only two phases in the game, namely the defect phase which exists for $0<\gamma<\gamma_0$ and quantum phase, which exists for $\gamma_0<\gamma<\pi/2$, where $\gamma_0$ is the point of phase transition at $T \to 0$ has been defined in Eq.~(\ref{gamma0}), an example for which has been plotted in Fig.~\ref{zerotemp}(c). $\gamma_0$ is the point of phase transition only at $T \to 0$ since magnetization becomes discontinuous only at $T \to 0$, which is the hallmark of a phase transition\cite{Baxter1982,Jaeger1998}. For all other finite temperatures, $\gamma_0$ marks the point of transition where the dominant strategy of the game changes. 
\subsubsection{Analogy with type-II superconductors}
At $T \to 0$, we observe two second order phase transitions for the QPD, this is similar to what is seen in type-II superconductors below the critical temperature $T_c$\cite{tinkham_1996,kittel_2005}. The quantum phase in the QPD game existing between $\gamma > \gamma_2$ can be compared with the Meissner phase of the type-II superconductor which is formed for external fields $H < H_{c1}$. The vortex phase of the superconductors between $H_{c1} < H < H_{c2}$ can be compared with the random phase of the QPD game between $\gamma_1 < \gamma < \gamma_2$. The normal phase of the type-II superconductor existing for $ H > H_{c2}$ can be compared to the classical phase of QPD existing between $0 < \gamma < \gamma_1$. 
The condition for the vortex phase in type-II superconductor to exist is for the Ginzburg-Landau parameter $\kappa > \frac{1}{\sqrt{2}}$, while the condition on the payoffs for the random phase of QPD to exist is $p+r<s+t$.  If the condition for the presence of the random phase in QPD is not maintained, then we find that there are two phases in the game with a phase transition at $\gamma_0$, which is similar to the phase transition of a type-I superconductor at $T_c$ in zero external field ($H=0$). But unlike the magnetic susceptibility $\chi = -1/4\pi$ in the Meissner phase, we find that the susceptibilities of the QPD game are zero in the quantum phase. 
\section{Conclusion}
In this paper, we have analyzed thermodynamic susceptibility in context of infinite player social dilemmas like Prisoner's Dilemma, Hawk-Dove game and quantum prisoner's dilemma. The infinite player game is constructed by mapping two-player game to spin-1/2 Ising model and then defining thermodynamic functions for infinite player game analogous to Ising model. 
The susceptibility for game is a effectively a measure of the turnover of players from one strategy to other in a game. The susceptibilities in context of game provide us with net change in player strategies, with the sign of susceptibility suggesting strategy players prefer to switch to. 

In Prisoner's Dilemma, we find that the turnover of players switching to cooperate strategy is positive for sucker's payoff and temptation, even though majority of players choose to defect. The turnover in the context of reward susceptibility is highly dependent on the game temperature($T$), while for punishment susceptibility, we find the net turnover of players preferring to switch to defect is dominant. 
In Hawk-Dove game, we find that the net turnover of players switching to Hawk is positive, in response to change in resource value, while turnover of players changing to Hawk is negative as cost of injury increases, implying that players prefer to switch to Dove strategy.

In quantum Prisoner's Dilemma game, we compare quantum to defect strategy. At finite game temperature ($T\neq 0$), we find that the net turnover of players switching to quantum strategy is positive in response to reward and punishment at higher entanglement. But at maximal entanglement, punishment and reward both do not influence players as corresponding susceptibilities are zero for low $T$. On the other hand, net turnover of players switching to quantum strategy is positive for sucker's payoff and temptation at lower entanglement. 
Increasing entanglement, causes temptation and sucker's susceptibility to become negative, implying that at higher entanglements, players prefer to switch to defect strategy. At maximal entanglement, we find that sucker's and temptation susceptibility becomes zero at low $T$, implying that entanglement inhibits players from switching their strategies. 

In the $T \to 0$ limit, we find that QPD has two second-order transitions, namely from all defect to random selection of strategies and from random selection to all quantum. This change in the Nash equilibrium for the game in response to entanglement is marked by a divergence in the susceptibilities at those transitions. We finally show the analogy of this behavior with the behavior seen in type-II superconductors.
Finally what is the main take home message of our work? 
The main message is two fold: one the susceptibility gives microscopic change in behavior of the players as compared to the macroscopic behavior understood from the magnetization. Second to identify and  characterize phase transitions in the game. 

We introduced the idea of susceptibility for the games in an attempt to better understand if there is any possibility of a phase transition in the game. We have tried to illustrate by means of examples, in fact we can observe telltale signs of such phase transitions in the Quantum Prisoner's Dilemma. Entanglement in Quantum Prisoner’s Dilemma (QPD) has a non-trivial role in determining the behavior of thermodynamic susceptibility. In the zero-temperature limit, we find that there are two second-order phase transitions in the quantum prisoner's dilemma game, marked by a divergence in the susceptibility. This behavior is similar to that seen in Type-II superconductors wherein also two second-order phase transitions are seen.
\section{Data Availability}
Data available in article-The data that supports the findings of this study are available within the article.
\acknowledgments
This work was supported by the grants- 1. ”Josephson junctions with strained Dirac materials and their application in quantum information processing” from SCIENCE \& ENGINEERING RESEARCH BOARD, New Delhi,
Government of India, Grant No. CRG/20l9/006258, Principal Investigator: Dr. Colin Benjamin, National Institute of Science Education and Research, Bhubaneswar, India, and 2. “Nash equilibrium versus Pareto optimality in N-Player games”, SERB MATRICS Grant No. MTR/2018/000070, Principal Investigator: Dr. Colin
Benjamin, National Institute of Science Education and Research, Bhubaneswar, India.
\bibliography{references.bib}
\bibliographystyle{apsrev4-2.bst}
 \end{document}